\documentclass[aps,prb,twocolumn,superscriptaddress,longbibliography]{revtex4-2}
\usepackage[utf8]{inputenc}
\usepackage{amsmath}
\usepackage{comment}
\usepackage{mathtools}
\usepackage{amssymb}
\usepackage{bm}
\usepackage{braket}
\usepackage{color}
\usepackage[varg]{txfonts}
\usepackage{varwidth}
\usepackage{dcolumn}
\usepackage[breaklinks,colorlinks=true,linkcolor=blue,urlcolor=cyan,citecolor=blue]{hyperref}
\usepackage{graphicx}

\newcommand{\CRPS}{CeRh$_{0.9}$Pd$_{0.1}$Sn}

\begin{document}
\title{Quantum Griffiths phase in the kagome Kondo lattice CeRh$_{0.9}$Pd$_{0.1}$Sn}

\author{Nan Tang}
\email{nan.tang@uni-a.de}
\affiliation{Experimental Physics VI, Center for Electronic Correlations and Magnetism, University of Augsburg, Augsburg 86159, Germany}

\author{Rajesh Tripathi}
\affiliation{ISIS Neutron and Muon Source, STFC, Rutherford Appleton Laboratory, Chilton, Oxon OX11 0QX, United Kingdom}

\author{Yasuyuki Shimura}
\affiliation{Department of Quantum Matter, Graduate School of Advanced Science and Engineering,
Hiroshima University, Higashi-Hiroshima 739-8530, Japan}

\author{Toshiro Takabatake}
\affiliation{Department of Quantum Matter, Graduate School of Advanced Science and Engineering,
Hiroshima University, Higashi-Hiroshima 739-8530, Japan}

\author{Devashibhai A. Adroja}
\affiliation{ISIS Neutron and Muon Source, STFC, Rutherford Appleton Laboratory, Chilton, Oxon OX11 0QX, United Kingdom}
\affiliation{Highly Correlated Matter Research Group, Physics Department, University of Johannesburg, Auckland Park 2006, South Africa}

\author{Philipp Gegenwart}
\email{philipp.gegenwart@uni-a.de}
\affiliation{Experimental Physics VI, Center for Electronic Correlations and Magnetism, University of Augsburg, Augsburg 86159, Germany}

\begin{abstract}
CeRhSn is a valence fluctuating heavy-fermion metal with a twisted Ce-kagome lattice, displaying zero-field quantum criticality, previously associated with geometrical frustration. The partial substitution of Rh by Pd in CeRh$_{1-x}$Pd$_x$Sn enlarges the unit-cell volume, suppresses valence fluctuations, decreases the Kondo temperature and stabilizes a possible long-range antiferromagnetic (AFM) ordered ground state with $T_N=0.8$~K at $x=0.5$. Previous thermodynamic and spectroscopic measurements for $x=0.1$ suggested a quantum critical spin liquid. We report low-temperature dilatometry and magnetization measurements on CeRh$_{0.9}$Pd$_{0.1}$Sn and compare with published low-$T$ specific heat data. The absence of a Gr\"uneisen parameter divergence excludes a conventional quantum criticality scenario. Instead, the weak power-law divergences signal non-Fermi liquid (NFL) behavior, according to a disorder-driven quantum Griffiths phase scenario. At low temperatures, a negative thermal expansion is found in fields above approximately 0.5~T and the NFL scaling breaks down, probably due to the polarization of AFM correlations.

\end{abstract}

\date{\today}
\maketitle

Heavy-fermion metals, where localized 4$f$ moments on rare-earth ions are embedded in a sea of conduction electrons, exemplify systems in which strong electron correlations give rise to emergent phenomena beyond conventional Fermi-liquid theory~\cite{Stewart_1984}. The key physics of the Kondo lattice arises from the competition between the Ruderman–Kittel–Kasuya–Yosida (RKKY) interaction, which promotes magnetic ordering, and the Kondo effect, which screens the local moments and favors a nonmagnetic, heavy Fermi-liquid ground state. This interplay makes Kondo lattices a central platform for investigating quantum criticality and novel phases such as unconventional superconductivity \cite{Lohneysen, Gegenwart_2008, Si_2010, Paschen_2021}. Further complexity may arise from valence fluctuations, magnetic frustration and/or structural randomness. For example, valence fluctuations lead to a drastic enhancement of superconductivity in pressurized CeCu$_2$Si$_2$ \cite{Yuan_2003, Seyfarth_2012}, while geometrical frustration in CePdAl leads to partial magnetic order \cite{Donni_1996, Oyamada_2008} and a pressure-tuned quantum critical phase \cite{Zhao_2019}. Structural randomness can lead to a significant smearing of quantum phase transitions~\cite{Lohneysen,Brando}. For instance, in the ferromagnetic CePd$_{1-x}$Rh$_x$, structural randomness leads to a unique Kondo-cluster-glass behavior \cite{Westerkamp_2009}.


In this Letter, we focus on \CRPS{} \cite{Yang_2017, Sundermann_2021, Tripathi_2022}, a heavy-fermion metal located near an antiferromagnetic (AFM) quantum critical point (QCP). This system presents a rare combination of three intertwined complexities: strong valence fluctuations, geometrical frustration, and structural randomness. Based on our experimental findings, we propose that these combined effects may give rise to a distinct form of quantum Griffiths phase behavior, whose properties arise from rare magnetically ordered regions embedded in a disordered phase near the QCP~\cite{CastroNeto_1998}.


CeRhSn crystallizes in the hexagonal ZrNiAl structure with 4$f$ atoms on a twisted kagome lattice in the basal plane, giving rise to geometrical frustration \cite{Donni_1996, Oyamada_2008, Tokiwa_2013, Lucas_2017, Shimura_2021, Zhao_2020, Zhao_2024}. Combined with a high Kondo temperature $T_{\rm K} \sim$ 200 K \cite{Shimada_2006} and strong valence fluctuations \cite{Slebarski_2002}, long-range magnetic order is fully suppressed down to the lowest temperatures in CeRhSn \cite{Kim_2003}. Despite this, CeRhSn retains Ising-like spin anisotropy at low temperatures, suggesting the presence of localized spin degrees of freedom and resulting in pronounced magnetic anisotropy \cite{Kim_2003, Kittaka_2021}.

Previous thermodynamic measurements down to very low temperatures revealed a pronounced anisotropy of the thermal expansion coefficient $\alpha/T$ below 1~K: while $\alpha/T$ diverges within the $ab$ plane, it remains constant along the $c$ direction~\cite{Tokiwa_2015}. 
The associated zero-field QCP, concluded from the divergence of the Grüneisen parameter, therefore appears to be insensitive to $c$-axis strain -- consistent with a scenario in which quantum criticality is driven by geometrical frustration, which is not affected by strain along the $c$ axis. 
Indeed, subsequent uniaxial-stress dilatometry on CeRhSn found a transition from quantum criticality to magnetic order under compressive uniaxial stress along the $a$ direction \cite{Kuechler_2017}. Moreover, a similar negative in-plane thermal expansion signature below 1~K, indicative of magnetic correlations, has also been found in the isostructural sister material CeIrSn, which even has twice as large Kondo temperature compared to CeRhSn \cite{Shimura_2021}. Finally, highly anisotropic quantum critical scaling has been found for both CeRhSn and CeIrSn at low temperatures in the rotational Gr\"uneisen parameter, which measures the relative temperature change with magnetic field rotation under adiabatic conditions~\cite{Yuasa_2025}.



For the substitution series CeRh$_{1-x}$Pd$_x$Sn, a monotonic increase in unit cell volume has been observed with increasing Pd content, along with the emergence of an AFM ordered ground state for higher $x$. Pronounced NFL behavior at $x = 0.1$ has been interpreted as a signature of an AFM QCP~\cite{Yang_2017}. Subsequent investigation by muon spin relaxation and inelastic neutron scattering revealed microscopic evidence for dynamical spin fluctuations at $x=0.1$, resembling a metallic spin-liquid close to the QCP~\cite{Tripathi_2022}. Thus, it is interesting to further explore CeRh$_{0.9}$Pd$_{0.1}$Sn by suitable physical properties at very low temperatures. For this purpose, we used the same single crystals investigated earlier and also reanalyzed the specific heat data from~\cite{Tripathi_2022}.

Linear thermal expansion and magnetostriction measurements were performed using a miniaturized high-resolution capacitive dilatometer \cite{Kuechler_2017_methods} adapted to a dilution refrigerator. We used a rectangular- shaped single crystal from the earlier study~\cite{Tripathi_2022} with lengths $L_0$ of 1.64 and 1.36 mm along the $a$ and $c$ directions, respectively, at room temperature. The linear thermal expansion coefficient
$\alpha(T)=L_0^{-1} d \Delta L(T)/dT$ is determined from the temperature derivative of the length change normalized by the sample length $L_0$, while the volume expansion coefficient $\beta(T)=V_0^{-1} d\Delta V/dT$ is obtained from $\beta=2\alpha_a + \alpha_c$.
Additionally, magnetization measurements using a superconducting quantum interference device (MPMS, Quantum Design), equipped with the He$^3$ option, were performed down to 0.4 K.



\begin{figure}[t]
\centering
\includegraphics[width=0.9\linewidth]{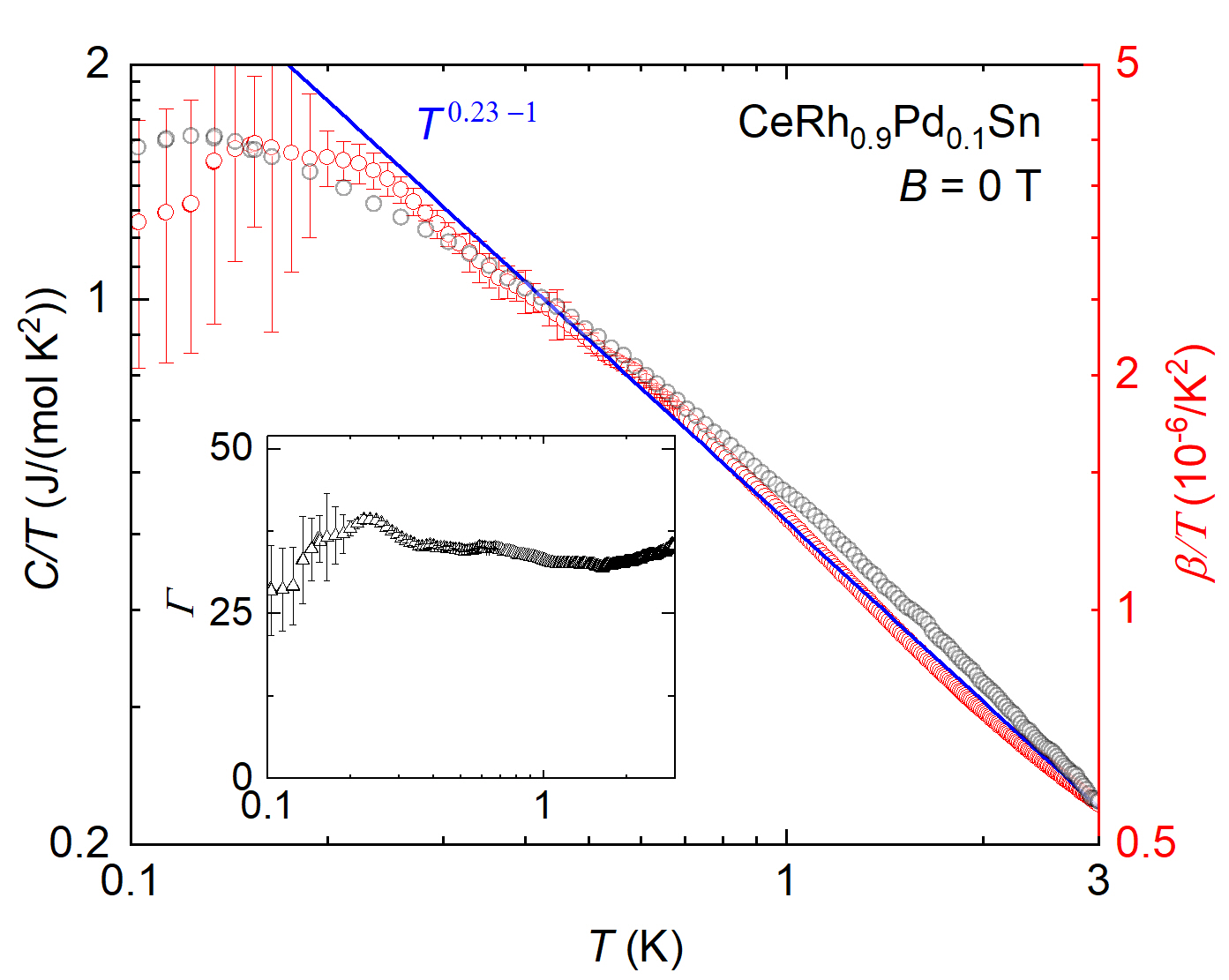}
\caption{Double-logarithmic plot of the temperature dependence of the magnetic specific heat divided by temperature~\cite{Tripathi_2022}, $C/T$ (left axis), and the volume thermal expansion coefficient divided by temperature, $\beta/T$ (right axis), of \CRPS{} single crystal in zero magnetic field. Open symbols represent the experimental data with error bars. The solid blue line indicates a $T^{\lambda-1}$ divergence with $\lambda=0.23$. The inset displays the temperature dependence of the Gr\"uneisen parameter $\Gamma$.} 
\label{fig1}
\end{figure}

\begin{figure}[t]
\centering
\includegraphics[width=1\linewidth]{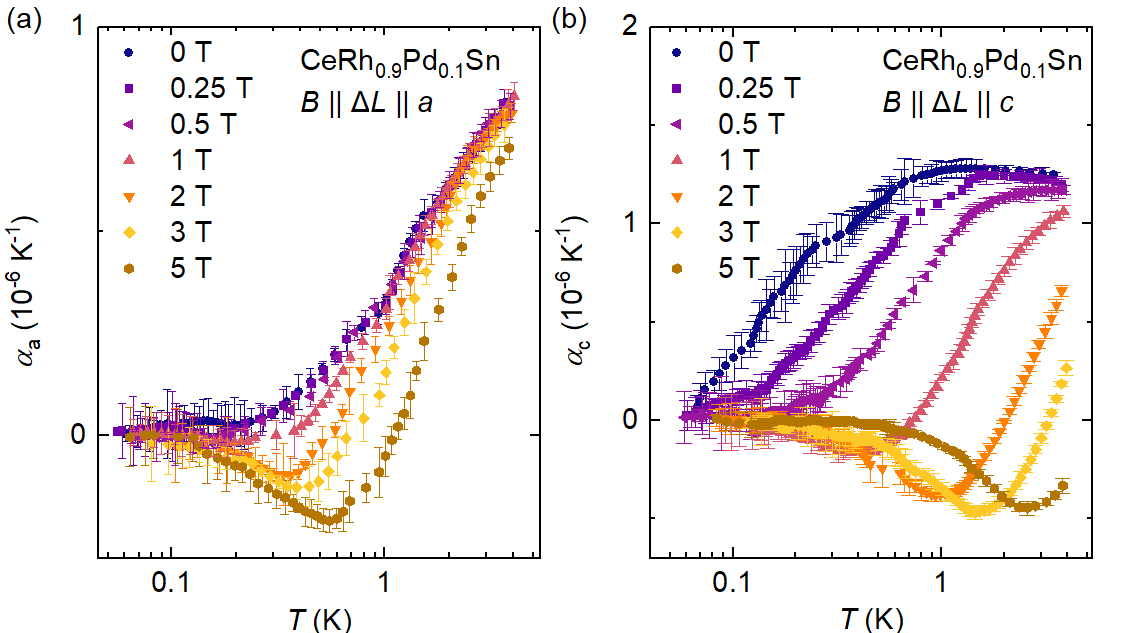}
\caption{Temperature dependence of the longitudinal thermal expansion coefficient (a) $\alpha_a$ and (b) $\alpha_c$ of \CRPS{} measured along the $a$-axis and $c$-axis, respectively, under various applied magnetic fields. The temperature axis is shown on a logarithmic scale to highlight low-temperature behavior.}
\label{fig2}
\end{figure}

Figure~\ref{fig1} compares the specific heat (after phonon subtraction) divided by temperature, $C(T)/T$ (adapted from Ref.~\cite{Tripathi_2022}) and volume thermal expansion coefficient divided by temperature, $\beta(T)/T$, of \CRPS{} in zero magnetic field, on double log scales with similar y-axis ranges of one decade. Note that the respective linear thermal expansion coefficients $\alpha_a$ and $\alpha_c$ are displayed in Fig.~\ref{fig2} and will be discussed later. While for a Fermi liquid, constant coefficients $C/T$ and $\beta/T$ are expected, the strong divergence found in both properties below 3 K signals pronounced NFL behavior. While broad maxima in $C/T$ and the ac susceptibility near 0.16~K indicated the formation of short-range magnetic correlations, the zero-field muon spin relaxation suggested persistent slow spin fluctuations down to 50 mK~\cite{Tripathi_2022}.

In case of a pressure-sensitive QCP, universal scaling of the critical contributions to the free energy results in the expectation of an algebraic divergence of the Gr\"uneisen parameter $\Gamma=B_{\rm M}V_{\rm m}\beta/C$, upon cooling~\cite{Zhu_2003,Kuechler_2003}. Here, the constants $B_{\rm M}$ and $V_m$ denote the bulk modulus and molar volume, respectively, with the values adapted from Ref.~\cite{Tokiwa_2015}. The inset of Fig.~\ref{fig1} displays $\Gamma(T)$, calculated from the data of the main panel, which is clearly non-divergent upon cooling from 3~K, excluding a pressure sensitive QCP. In fact, the Gr\"uneisen parameter shows only a weak temperature dependence. This is clearly reflected in the nearly parallel temperature trends of $C/T$ and $\beta/T$ shown in the main panel. Such behavior contrasts sharply with that of prototypical quantum critical materials~\cite{Gegenwart_2016}, where a divergent Gr\"uneisen parameter and its associated critical exponent are key indicators of the nature of quantum critical fluctuations. Furthermore, it is also different to the case of unsubstituted CeRhSn, where a low-temperature divergence of $\Gamma(T)$ has been reported~\cite{Tokiwa_2015}. On the other hand, the absence of a divergent Gr\"uneisen parameter at the smeared ferromagnetic quantum phase transition in CePd$_{1-x}$Rh$_x$ ($x \geq 0.65$) has been associated with a ‘‘Kondo-cluster-glass’’ state. In this scenario, magnetic clusters arise from regions with locally suppressed Kondo temperatures embedded within a nonmagnetic, valence fluctuating regions~\cite{Westerkamp_2009}. Given the overall valence fluctuating character of CeRhSn, the strong sensitivity of its AFM correlations to low Pd substitution, and the statistical disorder introduced by Rh-Pd mixing, it is tempting to explore similar mechanism for \CRPS{}. Indeed, we can describe both $C(T)/T$ and $\beta(T)/T$ by a weak power law divergence over roughly one decade in temperature (cf. the blue line in Fig.~\ref{fig1}). Such behavior is expected within the quantum Griffiths phase scenario for disordered Kondo lattices close to quantum criticality~\cite{CastroNeto_1998, CastroNeto_2000}. It deals with quantum fluctuations of magnetic clusters, giving rise to singular contributions to the temperature dependencies of various physical properties in a finite region, defined as the quantum Griffiths phase, causing weak power-law singularities in specific heat, $C/T \sim T^{{\rm \lambda}-1}$, magnetic susceptibility, $\chi \sim T^{\lambda -1}$ and isothermal magnetization $M\sim B^{\rm \lambda}$ with $0 \leq {\rm \lambda} \leq 1$. As shown by the blue line in Fig.~\ref{fig1} a weak power-law divergence with ${\lambda} =0.23$ can approximate the coefficients of specific heat and thermal expansion. Interestingly, a classical Griffiths phase has recently been proposed for the elevated-temperature behavior (between 6 and 220 K) of polycrystalline CeRhSn~\cite{Slebarzki_2025}. 

To further examine possible quantum Griffiths-phase behaviour in \CRPS, we performed thermal expansion measurements under magnetic fields applied along the $a$ and $c$ directions down to 60 mK, as shown in Fig.~\ref{fig2}. In zero field, the overall thermal expansion is smaller for $\Delta L \parallel a$ compared to $\Delta L \parallel c$. This anisotropy is consistent with the behavior observed in undoped CeRhSn, where $\alpha_c > \alpha_a$ for temperatures above 0.2 K \cite{Tokiwa_2015}. The light doping of 10\% Pd does not appear to alter this anisotropic relationship. With increasing magnetic field, both $\alpha_a$ and $\alpha_c$ begin to develop a dip-like feature, which shifts towards higher temperatures with increasing field. This shift is stronger for $B\parallel c$ and accompanying magnetostriction measurements shown in the supplemental material~\cite{SuppMat} confirm that $B\parallel c$ is the magnetic easy direction.

\begin{figure}[t]
\centering
\includegraphics[width=1\linewidth]{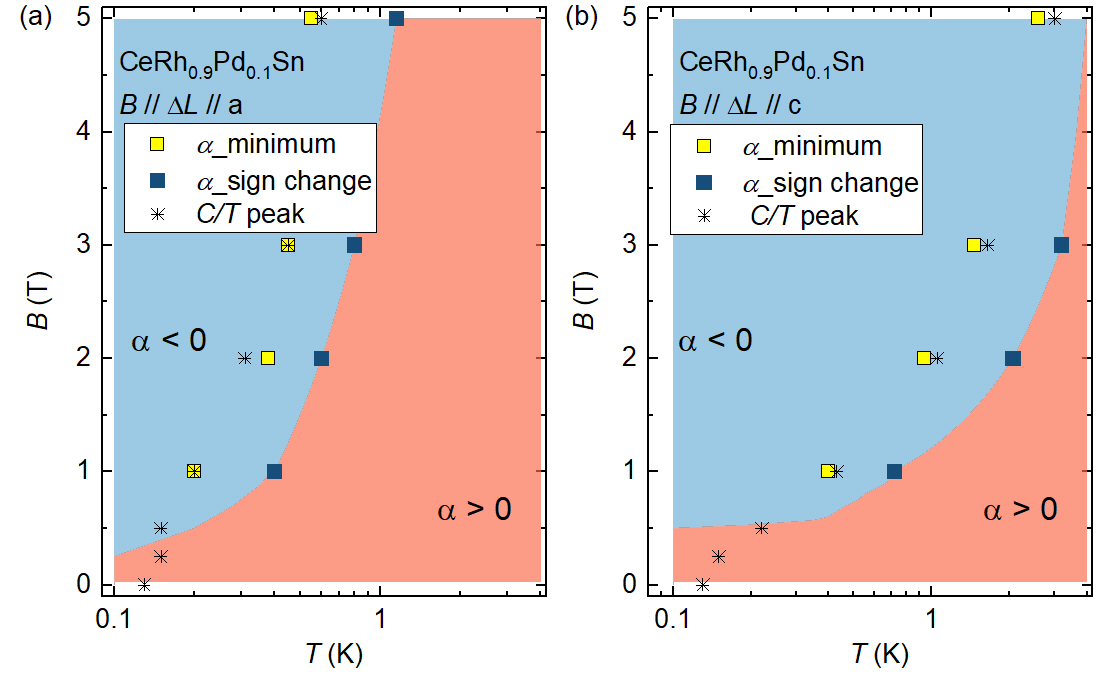}
\caption{$B-T$ phase diagram of \CRPS~for $B\parallel a$ (a) and $B\parallel c$ (b). Yellow and blue squares mark the positions of the minima in $\alpha (T)$ and the sign changes defined by $\alpha =0$, respectively. Asterisks indicate the peak positions extracted from $C/T$ data from Ref. \cite{Tripathi_2022}. Regions with positive/negative $\alpha(T)$ are shaded in red/blue.}

\label{fig3}
\end{figure}

Since the thermal expansion coefficient $\alpha$ is proportional to the negative pressure derivative of the entropy  ($\alpha \propto -\partial S/\partial p$), constructing a phase diagram based on the sign of $\alpha$ and the location of its minima provides valuable insight into the underlying entropy landscape, as shown in Fig.~\ref{fig3}. In the phase diagrams, the line where $\alpha = 0$ serves to separate regions with opposite signs of $dS/dp$, highlighting changes in the thermodynamic response of the system.

According to the Gr\"uneisen relation, the sign of $\alpha$ reflects the pressure dependence of the system's dominating energy scale~\cite{Gegenwart_2016}. In Ce-based Kondo systems, $\alpha>0$ is typically observed when the Kondo effect dominates over magnetic intersite coupling, as the Kondo temperature generally increases under pressure, due to strengthened $c-f$ hybridization, reflecting a decrease in magnetic entropy. Conversely, magnetic order in Ce systems is suppressed by pressure, and the corresponding increase in entropy thus leads to a negative $\alpha$. Often, even the presence of quasi-static magnetic correlations leads to $\alpha<0$, as reported for CeRhSn under uniaxial pressure \cite{Kuechler_2017} and CeIrSn at ambient pressure at low temperatures \cite{Shimura_2021}. In \CRPS, the observed negative $\alpha$ indicates that (quasi-static) short-range magnetic order is destabilized by pressure. Interestingly, the temperature positions of the minima in $\alpha(T)$ roughly coincide for fields above 0.5~T with those of the maxima in $C/T$ in both phase diagrams in Fig.~\ref{fig3}. This suggests a common origin, possibly related to the Zeeman splitting of the Ce Kramers doublets~\cite{Tripathi_2022} and/or field-polarization of rare magnetic clusters.


To further investigate the low-temperature properties of \CRPS, we measured the magnetization $M$ down to 0.4 K with the magnetic field applied along both crystallographic directions. As shown in Fig. 4(a), the magnetic susceptibility $\chi=M/B$ exhibits power-law behavior of the form $\chi\sim T^{\lambda-1}$, with $\lambda = 0.48$ for $B \parallel a$ and $\lambda = 0.37$ for $B \parallel c$, respectively. Fig. 4(b) shows the field dependence of the magnetization. The magnetization $M(B)$ for $B \parallel c$ is much larger than for $B \parallel a$, preserving the anisotropy observed in undoped CeRhSn, where the $c$ axis is the easy axis. Thermal expansion at $B>0$ shown in Fig. 2 also indicates that $c$ is the magnetic easy direction. Moreover, the magnetization data in Fig. 4(b) reveal a scaling behavior of the form $M/B = T^{-\alpha}g(B/T^{\beta})$. The data collapse on a single universal curve with scaling exponents $\alpha = 0.47$, $\beta = 0.88$ for $B \parallel a$, and $\alpha = 0.39$, $\beta = 0.6$ for $B \parallel c$, as shown in the insets. Notably, the extracted values of $\alpha$ are in good agreement with the power-law exponent $\lambda = 0.48$ and 0.37 obtained from $\chi (T)$, reinforcing the consistency of the quantum Griffiths phase scenario in \CRPS.


\begin{figure}[t]
\centering
\includegraphics[width=0.8\linewidth]{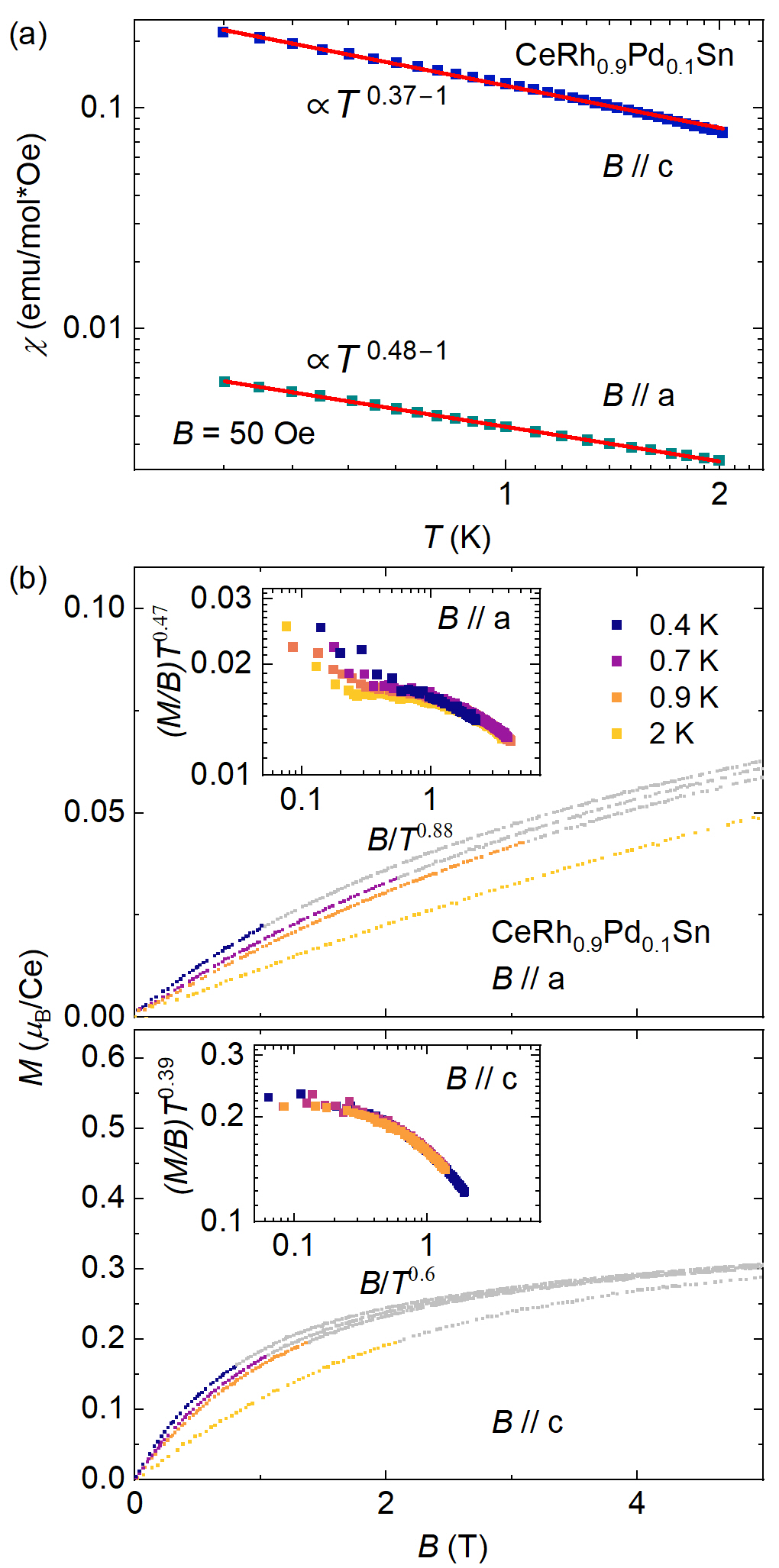}
\caption{(a) Temperature dependence of magnetic susceptibility $\chi$ of \CRPS{} for both field orientations $B \parallel a$ and $B \parallel c$, plotted on a double-logarithmic scale. Squares represent experimental data, and solid red lines indicate power-law fits of the form  $\chi(T) \propto T^{\lambda -1}$. (b) Field dependence of magnetization $M$ for both field directions. Colored data points correspond to the temperature regime where the thermal expansion $\alpha > 0$ are included in the scaling analysis; gray points ($\alpha < 0$ ) are excluded. Insets show scaling plots of $(M/B)T^{\alpha}$ vs $g (B/T^{\beta})$ scaling in the temperature range $0.4< T< 2$ K.}

\label{fig4}
\end{figure}



We now consider the possible origin of the quantum Griffiths phase behavior in \CRPS. Photoelectron spectroscopy on CeRh$_{1-x}$Pd$_x$Sn with hard x-rays revealed that the $c-f$ hybridization is strongly suppressed by the Pd doping ratio $x$, driving the system from the strongly intermediate valence regime, where,
usually, magnetic order does not occur, into the AFM regime with low Kondo temperature~\cite{Sundermann_2021}. Although the $x=0.1$ compound remains in the intermediate-valence regime on average, the statistical distribution of Pd atoms introduces local chemical disorder, which is likely to give rise to a broad distribution of the local Ce valence states. This distribution likely includes a tail toward nearly trivalent Ce$^{3+}$ configurations with low Kondo temperatures, giving rise to rare regions with localized magnetic moments. On the other hand, non-substituted intermediate valent CeRhSn displays anisotropic zero-field quantum criticality, and local moment metamagnetism, which was interpreted as signature of geometrical frustration leading to an incomplete Kondo screening of the Ce moments~\cite{Tokiwa_2015}. This suggests that NFL behavior near $x=0.1$ in CeRh$_{1-x}$Pd$_x$Sn~\cite{Yang_2017} is governed by magnetic frustration as well, as concluded earlier~\cite{Tripathi_2022}. Here we propose that it is the combination of geometrical frustration, opposing Kondo singlet formation, with structural disorder and valence fluctuations that may account for the unusual properties in CeRh$_{1-x}$Pd$_x$Sn for low $x$. 

It is worth to recap the temperature dependence of the electrical resistivity $\rho(T)$ in these materials~\cite{Yang_2017}, which reveals the significant structural disorder, as indicated by large residual resistivity values $\sim 100~\mu\Omega$cm, particularly for in-plane current directions. While unsubstituted CeRhSn exhibits only a weak signature of the onset of coherent Kondo-lattice behavior below 100~K, the resistivity of \CRPS{} continues to increase upon cooling, down to the lowest measured temperatures, indicating a complete absence of Kondo-lattice coherence. This reflects a substantial degree of structural disorder, which likely promotes the formation and stabilization of rare magnetic regions with increasing Pd content $x$, extending their influence down to the milli-Kelvin regime. Since the $x=0.1$ material remains, on average, in the valence-fluctuating regime~\cite{Sundermann_2021}, the distribution of local $T_K$ values is expected to be significantly broader than in unsubstituted CeRhSn. In combination with the incomplete formation of Kondo singlets due to geometrical frustration, this may lead to the emergence of rare magnetic regions whose slow dynamics gives rise to the observed weak power-law behavior in magnetic and thermodynamic properties. Furthermore, the persistent spin dynamics observed down to at least 80 mK evident from $\mu$SR and inelastic neutron scattering~\cite{Tripathi_2022} would be compatible with a universal disorder-driven spin liquid, as theoretically proposed in~\cite{Miranda_2005}.
While quantum Griffiths phases are typically associated with {\it ferromagnetic} rare regions \cite{Vojta_2010}, CeRhSn and its Pd-doped variants display AFM correlations and spin-liquid behavior, making this an unusual and intriguing realization. Notably, long-range AFM order does not emerge until the Pd concentration exceeds approximately 65\%, as evidenced by the full recovery of the magnetic ground-state doublet entropy in specific heat measurements and the appearance of spin-flop–like behavior in $M (B)$ curves \cite{Yang_2017}. 

Finally, we note that recent studies on polycrystalline, unsubstituted CeRhSn have revealed nanometer-sized crystalline grains separated by intercrystalline regions containing point defects and dislocations~\cite{Slebarzki_2025}.
Magnetic disorder signatures, observed in the temperature dependence of both the magnetic susceptibility and specific heat in these samples, were interpreted within a classical Griffiths phase framework below approximately 220 K. In this regime, weakly coupled rare regions of ordered magnetic clusters can be easily polarized by small magnetic fields, with a possible crossover to a quantum Griffiths phase emerging around 6 K.
In contrast, such behavior is absent in impurity-free single crystals~\cite{Kim_2003}, which below 1 K display a clear divergence of the Gr\"uneisen parameter, providing strong evidence for intrinsic quantum criticality in the absence of disorder-driven Griffiths phase physics~\cite{Tokiwa_2015}.

In summary, motivated by the observation of an AFM quantum phase transition in the substitution series CeRh$_{1-x}$Pd$_x$Sn and pronounced NFL behavior at $x=0.1$~\cite{Yang_2017,Tripathi_2022}, we studied the low-temperature thermal expansion and Gr\"uneisen parameter of $x=0.1$ single crystals. 
Strikingly, we find a weak power-law divergence in the thermal expansion coefficient $\alpha/T$, mirroring that in the specific heat coefficient $C/T$, which results in an almost temperature independent Gr\"uneisen parameter. This behavior excludes the possibility of a pressure-sensitive QCP. 
Similar weak power-law divergences also appear in the temperature dependence of the low-field magnetic susceptibility and in the initial field dependence of the magnetization at low temperatures.

Notably, the scaling behavior breaks down above approximately 0.5 T, where a crossover to negative thermal expansion is observed, potentially signaling the polarization of interacting magnetic clusters.
Previous studies have shown a smooth increase of the unit-cell volume accompanied by a pronounced suppression of $c–f$ hybridization and valence fluctuations with increasing Pd substitution $x$ in \CRPS~\cite{Sundermann_2021}.
This likely gives rise to a broad distribution of local Kondo temperatures, including a low $T_K$ tail. Such distribution facilitates the formation of rare regions with interacting magnetic moments, a key prerequisite for the quantum Griffiths phase scenario. 
Earlier microscopic measurements revealed persistent spin-liquid dynamics down to lowest temperatures~\cite{Tripathi_2022}. Combined with the quantum Griffiths scaling in both thermodynamic and magnetic properties reported above, these results suggest that a disorder-induced, universal spin-liquid state provides a compelling explanation for the intriguing behavior of \CRPS.

\section*{ACKNOWLEDGMENTS}
This work was supported by the Deutsche Forschungsgemeinschaft (DFG, German Research Foundation) via TRR 360 (project No.~492547816), JSPS KAKENHI Grants-in-Aid for Scientific Research (Grants No.~JP17K05545). N.T. was supported by the Alexander von Humboldt Foundation. We appreciate T. Onimaru, Y. Tokiwa, and  A.M. Strydom for helpful discussions. 

\section*{DATA AVAILABILITY}
The data that support the findings of this article are openly available~\cite{Zenodo}.



\end{document}